\documentclass[pra,aps,superscriptaddress,footinbib,twocolumn,longbibliography]{revtex4-2}

\usepackage{mathrsfs}
\usepackage{amsfonts}
\usepackage{amssymb}
\usepackage{amsmath}
\usepackage{graphicx}
\usepackage[usenames,dvipsnames]{color}
\usepackage[colorlinks=true,citecolor=blue,linkcolor=magenta]{hyperref}
\usepackage{lmodern}
\usepackage{amsthm}
\usepackage{dsfont}
\usepackage{natbib}
\usepackage{bm}
\usepackage{tabularx}
\usepackage{booktabs}

\usepackage{algorithm}
\usepackage[noend]{algpseudocode}

\begin{document}

\title{Achieving High-Quality Portfolio Optimization with the Variational Quantum Eigensolver}

\author{Zhonggang Lv}
\affiliation{Future Science and Technology Research Lab, 
China Mobile (Suzhou) Software Technology Company Limited, Suzhou 215163, China}

\author{Zhenyuan Ma}
\affiliation{Future Science and Technology Research Lab, 
China Mobile (Suzhou) Software Technology Company Limited, Suzhou 215163, China}

\author{Binglei Wang}
\affiliation{Future Science and Technology Research Lab, 
China Mobile (Suzhou) Software Technology Company Limited, Suzhou 215163, China}

\author{Anbang Wang}
\email{wanganbang@cmss.chinamobile.com}
\affiliation{Future Science and Technology Research Lab, 
China Mobile (Suzhou) Software Technology Company Limited, Suzhou 215163, China}

\begin{abstract}
Portfolio optimization lies at the core of quantitative finance 
and aims to determine how assets should be allocated 
to balance expected returns against risk. 
It can be formulated as a Quadratic Unconstrained Binary Optimization (QUBO) problem, 
which is NP-hard. Quantum computing offers the potential 
to solve such problems more efficiently than classical methods.
In this work, we employ the Variational Quantum Eigensolver (VQE) 
to address the portfolio optimization problem. 
To increase the likelihood of converging to high-quality solutions, 
we propose using the Weighted Conditional Value-at-Risk (WCVaR) as the cost function 
and the Covariance Matrix Adaptation Evolution Strategy (CMA-ES) as the optimizer. 
Our experiments are conducted using both classical simulations and quantum hardware
on the Wuyue QuantumAI platform.
Together, these results demonstrate that the combination of WCVaR and CMA-ES 
improves the performance of VQE for portfolio optimization
and provides a practical route for applications on NISQ devices.
\end{abstract}

\maketitle

\section{Introduction}

Portfolio optimization~\cite{Markowitz1952PortfolioSelection} is a central task in finance 
concerned with allocating capital across different assets. 
It plays a crucial role in modern investment management 
by helping investors construct efficient portfolios and 
make informed decisions in complex financial markets. 
In many practically relevant settings, this problem can be formulated 
as a QUBO problem~\cite{Chang2000CardinalityPortfolio,Palmer2022CardinalityQuantum}, 
which is generally NP-hard and becomes computationally challenging 
for large-scale instances on classical computers.
Quantum computing offers the potential to solve such QUBO problems more efficiently, 
paving the way for significant advancements in portfolio optimization~\cite{Gemeinhardt2024NISQMapping}. 
Among the quantum algorithms, the Variational Quantum Eigensolver~\cite{peruzzo2014,
    Buonaiuto2023BestPracticesPortfolioOptimization,le_variational_2025}
and the Quantum Approximate Optimization Algorithm (QAOA)~\cite{
    Edward2014QuantumApproximateOptimizationAlgorithm,Brandhofer2022BenchmarkingPerformancePO,nodar_scaling_2025} 
are two promising candidates 
particularly well-suited for execution 
on Noisy Intermediate-Scale Quantum (NISQ) devices~\cite{Preskill2018NISQ,Gemeinhardt2024NISQMapping}.

However, solving a QUBO problem is fundamentally different from 
finding the ground state of a typical chemical Hamiltonian. 
The Ising Hamiltonian associated with a QUBO problem is composed of mutually commuting terms, 
and its ground state, if non-degenerate, is simply a computational basis state. 
Therefore, the conventional expectation value of energy is not always the most suitable cost function. 
To address this issue, Barkoutsos et al.~\cite{Barkoutsos2020ImprovingVariational} 
proposed using Conditional Value-at-Risk (CVaR), 
a risk measure widely used in finance~\cite{Acerbi2002OnCoherence}, 
as the optimization objective. The CVaR objective is closely aligned with the practical goals 
of portfolio optimization and can improve the performance and robustness of VQE and QAOA.

Nevertheless, CVaR retains only a small fraction of measurement outcomes 
with the lowest energies when evaluating the objective. 
Although a smaller retained fraction often leads to better optimization behavior, 
it also discards a substantial amount of measurement data. 
This feature is unfavorable for real quantum hardware, 
where measurements are expensive and shot resources are limited.
To overcome this limitation, we propose a Weighted CVaR (WCVaR) cost function, 
which replaces hard truncation with an energy-dependent weighting scheme over sampled outcomes. 
By assigning larger weights to more favorable low-energy samples 
while still utilizing the full set of measurement results, 
WCVaR preserves the advantages of CVaR without wasting measurement data. 
This makes it particularly attractive for practical execution on real quantum devices.
In addition, we employ the Covariance Matrix Adaptation Evolution 
Strategy (CMA-ES)~\cite{Hansen2023CMAEvolutionStrategytutorial} 
as the classical optimizer to alleviate convergence difficulties caused 
by potentially non-smooth or ill-conditioned objective landscapes. 
Classical simulations show that the proposed combination of WCVaR and CMA-ES 
improves the optimization performance of VQE for portfolio optimization. 
Experiments on real quantum hardware further demonstrate the feasibility of the method 
and show a clear concentration of probability on the optimal solution.

The remainder of this paper is organized as follows. 
Section~\ref{sec:background} provides a brief introduction to the relevant background, 
including the modeling of the portfolio optimization problem based on historical closing-price data, 
its conversion into a QUBO problem, and its solution using the VQE algorithm. 
Section~\ref{sec:design} describes the detailed implementation of our VQE approach. 
Section~\ref{sec:results} presents the results from both numerical simulations 
and experiments on real quantum hardware. 
Finally, Section~\ref{sec:conclusions} concludes the paper.

\section{Background}
\label{sec:background}

Portfolio optimization aims to determine an asset allocation 
that achieves a favorable trade-off between expected return and risk. 
Modern Portfolio Theory (MPT), introduced by Harry Markowitz in 1952~\cite{Markowitz1952PortfolioSelection}, 
is a widely accepted framework for portfolio optimization. 
It addresses the trade-off between risk and return through mean-variance analysis, 
emphasizing diversification to achieve efficient portfolios. 
In this section, we demonstrate how to formulate the portfolio optimization problem 
given market data and how to reduce it to the 
QUBO problem and the corresponding Ising model.

\subsection{Portfolio Optimization}
\label{sec:po}
The market data are represented as a matrix $P$ of size $M \times N$, 
where $N$ is the number of assets, $M$ is the number of observation times, and 
each entry $P_{ki}$ denotes the price of asset $A_i$ at observation time $t_k$.
Let $T = (t_1, t_2,\ldots, t_M)^T$ be the vector of observation times.
We define the return of asset $A_i$ between consecutive observation times 
$t_{k-1}$ and $t_k$ as 
\begin{align}
r_{ki} &= \frac{P_{ki} - P_{k-1,i}}{P_{k-1,i}}.
\end{align}
Using these returns, we compute the expected return of asset $A_i$ and 
the covariance between asset $A_i$ and $A_j$ as
\begin{align}
\mu_i &= \frac{1}{M}\sum_{k=1}^M r_{ki}, \\  
\sigma_{ij} &= \frac{1}{M-1} \sum_{k=1}^M (r_{ki} - \mu_i)(r_{kj} - \mu_j).
\end{align}

Suppose we have a total budget $B$ that we wish to
allocate among $N$ assets in the portfolio. 
Let $b_i$ be the portion of the budget allocated to asset $A_i$. 
These allocations must satisfy the following constraint:
\begin{align}
\sum_i b_i &\le B.\label{eq:constraint1}
\end{align}

Portfolio optimization involves two primary objectives: 
maximizing return and minimizing risk. The expected return, 
computed from historical market data, serves as a reasonable estimate for future performance. 
Therefore, the first objective function to maximize is the expected return, given by:
\begin{align}
C_1(\boldsymbol b) &= \sum_{i=1}^N \mu_i b_i,
\end{align}
where $\boldsymbol{b} = (b_1, b_2,\ldots, b_N)^T$ is the vector of investment allocations.
The risk of the portfolio is typically measured by its volatility, 
which is the square root of the portfolio variance. 
Hence, the second objective function to minimize is the portfolio variance:
\begin{align}
C_2(\boldsymbol b) &= \sum_{ij} \sigma_{ij} b_i b_j.
\end{align}

The portfolio optimization problem is inherently a multi-objective optimization problem, 
involving two objective functions and one constraint. 
To solve such a problem, one seeks the Pareto front, also known as 
the efficient frontier in finance, which consists of solutions that cannot 
be improved in one objective without worsening the other. 
A common approach to finding the Pareto front is 
the scalarization method using linear weighting. In this method, 
the two objectives are combined into a single objective function 
by introducing a weighting factor $\lambda$:
\begin{align}
C'(\boldsymbol b) &= -\lambda C_1(\boldsymbol b) + (1-\lambda) C_2(\boldsymbol b).
\label{eq:costpb}
\end{align}
Our aim is to minimize the combined objective function $C'(\boldsymbol b)$ 
subject to the constraint \eqref{eq:constraint1}.

\subsection{Quadratic Unconstrained Binary Optimization}

To make the problem compatible with quantum optimization algorithms, 
we next reformulate it into a QUBO problem. 
Since the QUBO framework is defined over binary decision variables, 
the continuous portfolio weights must first be discretized, 
so that each portfolio configuration can be encoded by a binary string. 
Furthermore, the original constraint is absorbed into the objective function 
through a penalty term, thereby converting the constrained problem into an unconstrained one. 
With these treatments, the portfolio optimization problem can be cast into the standard QUBO form, 
which then allows a natural mapping to an Ising Hamiltonian for quantum computation.

QUBO is a combinatorial optimization problem which can be formulated as follows:
\begin{align}
Q(\boldsymbol x) &= \sum_{i<j} q_{ij}x_i x_j + \sum_i q_i x_i + c,
\label{eq:qubog}
\end{align}
where the variables $x_i$ are binary, and parameters $q_{ij}$, $q_i$ and $c$ are real numbers.

In general, the total budget $B$ and the individual investment amounts $b_i$ are real numbers. 
Discretizing them into binary variables can significantly increase 
the number of variables in the QUBO formulation. 
To avoid this overhead, we assume that the total budget $B$ is an integer 
and that each individual allocation $b_i$ is represented by a binary variable 
indicating whether asset $A_i$ is selected. Without loss of generality, 
we set $B=N/2$. In the following, we use $x_i\in\{0,1\}$ to represent 
the binary decision variable corresponding to asset $A_i$, 
ensuring consistency with the notation used in the QUBO problem~\eqref{eq:qubog}. 
Our goal then becomes minimizing the objective function
\begin{align}
C'(\boldsymbol x) &= -\lambda C_1(\boldsymbol x) + (1-\lambda) C_2(\boldsymbol x)
\label{eq:costpx}
\end{align}
subject to the constraint
\begin{align}
\sum_{i=1}^N x_i = N/2.
\label{eq:constraint2}
\end{align}

The constraint in Eq.~\eqref{eq:constraint2} can be incorporated into 
the objective function using a penalty method. Specifically, 
we add a penalty term of the form $p(\sum_{i=1}^N x_i -N/2)^2$, 
where $p>0$ is a sufficiently large penalty coefficient. 
The resulting final objective function to minimize is:
\begin{align}
C(\boldsymbol x) &= -\lambda C_1(\boldsymbol x) + (1-\lambda) C_2(\boldsymbol x) 
+ p\left(\sum_{i=1}^N x_i -\frac{N}{2}\right)^2 \\
&= -\lambda \sum_{i=1}^N \mu_i x_i + (1-\lambda) \sum_{i=1}^N\sum_{j=i+1}^N \sigma_{ij} 
x_i x_j \notag\\
&\phantom{=}+ p\left(\sum_{i=1}^N x_i -\frac{N}{2}\right)^2.
\label{eq:cost}
\end{align}

\subsection{Variational Quantum Eigensolver}

The QUBO problem is mathematically equivalent to the Ising model. 
To transform the problem, we introduce a spin vector $\boldsymbol s$,
where each element is defined as $s_i = 1-2x_i$, 
mapping the binary variables $x_i\in\{0, 1\}$ to spins $\{1, -1\}$. 
The objective function in Eq.~\eqref{eq:cost} can then be rewritten 
in the (classical) Ising form:
\begin{align}
E(\boldsymbol s) &= -\sum_{i=1} h_i s_i - \sum_{i<j} J_{ij} s_i s_j,
\end{align}

where $h_i$ and $J_{ij}$ are real coefficients derived 
from the QUBO objective function~\eqref{eq:cost}.
The corresponding quantum Ising Hamiltonian is obtained 
by replacing the classical spin variables $s_i$ with Pauli 
operators $\hat Z_i$:
\begin{align}
\hat H &= -\sum_{i=1} h_i \hat Z_i - \sum_{i<j} J_{ij} \hat Z_i \hat Z_j.
\end{align}

Like QUBO, the (classical) Ising model is generally NP-hard 
and therefore cannot be solved efficiently in the worst case. 
However, if we can find the ground state of the corresponding quantum Ising Hamiltonian, 
we can determine the ground state of the classical Ising model, thereby 
solving the original QUBO problem. Finding the ground state of a general quantum system 
is known to be QMA-complete, which is considered the quantum analog of NP-completeness. 
In principle, even quantum computers are not expected to solve such problems efficiently in general. 
Nevertheless, several heuristic quantum algorithms have been proposed 
that may offer advantages for certain problem instances. 
These include VQE, QAOA, and quantum annealing~\cite{Johnson2011QuantumAnnealingWithManufacturedSpins}. 
In this work, we focus on the VQE algorithm.

VQE is a hybrid quantum-classical algorithm. 
In this approach, a quantum computer is used to prepare a quantum state 
and measure the expectation value of a given Hamiltonian, 
while a classical computer performs an optimization to minimize this expectation value. 
Given a Hamiltonian $\hat H$, and a parameterized trial wave function (known as an ansatz) 
$\psi(\boldsymbol{\theta})$, the VQE algorithm seeks the optimal parameter vector $\boldsymbol\theta_m$ 
that minimizes the energy expectation value (the cost function), or Rayleigh quotient:
\begin{align}
E(\boldsymbol{\theta}) &= 
\frac{\langle \psi(\boldsymbol{\theta}) | \hat H | \psi(\boldsymbol{\theta}) \rangle}
{\langle \psi(\boldsymbol{\theta}) | \psi(\boldsymbol{\theta}) \rangle}. 
\label{eq:vqecost}
\end{align}
If we assume the state is normalized, the denominator equals one and can be omitted.

The VQE algorithm was originally proposed and is widely used to solve problems 
in physics and chemistry. The Ising Hamiltonian represents a special case, 
as its eigenstates are the computational basis states. 
Let the eigensystem of the Ising Hamiltonian be defined by
\begin{align}
\hat H |n\rangle &= E_n |n\rangle,
\end{align}
where the eigenstates are ordered in ascending order of energy, i.e.,
$|1\rangle$ denotes the ground state with energy $E_1$.
For a given parameter $\boldsymbol\theta^*$, the variational wave function 
can be expanded in this eigenbasis as
\begin{align}
|\psi(\boldsymbol\theta^*)\rangle &= \sum_n c_n(\boldsymbol\theta^*) |n\rangle, 
\label{eq:expansion}
\end{align}
and the corresponding energy expectation value is
\begin{align}
E(\boldsymbol\theta^*) &= \sum_n |c_n(\boldsymbol\theta^*)|^2 E_n.
\end{align}

We already know that the ground state is $|1\rangle$, with ground state energy $E_1$.
This state can, in principle, be reached in the expansion~\eqref{eq:expansion} 
by setting $c_1(\boldsymbol\theta^*) = 1$ 
and all other coefficients $c_i(\boldsymbol\theta^*)=0$.
However, it is generally difficult to design an ansatz capable of 
collapsing exactly to a specific computational basis state. 
Consequently, minimizing the energy expectation value~\eqref{eq:vqecost} 
may not be an effective strategy for guiding the VQE algorithm toward the true ground state.

Instead, the Conditional Value-at-Risk (CVaR) has been proposed 
as a more suitable cost function~\cite{Barkoutsos2020ImprovingVariational}.
Suppose we prepare a parameterized trial wave function $\psi(\boldsymbol\theta)$ 
and measure it in the computational basis, obtaining a set of bitstrings $\boldsymbol x_k$.
We then compute the energy of each measured bitstring as
$E_k = \langle \boldsymbol x_k|\hat H|\boldsymbol x_k\rangle$.
This process is repeated $K$ times, yielding a set of sampled energies.
These energies are sorted in ascending order to form the ordered sample set 
$\boldsymbol E = (E_{(1)}, E_{(2)}, \ldots, E_{(K)})$, 
where $E_{(k)}$ denotes the $k$-th smallest value. 
For a given confidence level $\alpha\in (0,1]$, CVaR is defined as 
the average of the smallest $\lceil \alpha K\rceil$ energies:
\begin{align}
\mathrm{CVaR}_{\alpha}(\boldsymbol E) &= 
\frac{1}{\lceil \alpha K\rceil}\sum_{k=1}^{\lceil \alpha K\rceil} E_{(k)}.
\label{eq:cvar}
\end{align}
When $\alpha=1$, $\mathrm{CVaR}_{1}$ reduces to the mean of the sampled energies
\begin{align}
\bar E &= \frac{1}{K}\sum_{k=1}^{K} E_k,
\end{align}
which approximates the standard energy expectation value~\eqref{eq:vqecost}.
When $\alpha$ is sufficiently small such that $\lceil \alpha K\rceil=1$, 
$\mathrm{CVaR}_{\alpha}$ becomes the minimum observed energy. 
If the variational ansatz has non-zero overlap with the true ground state 
for all parameter values, this minimum can, in principle, 
converge to the ground-state energy. However, for very small $\alpha$, 
the cost function becomes non-smooth, making it difficult for classical optimizers 
to navigate the landscape effectively. Moreover, in this regime, the CVaR objective 
retains only a tiny subset of the lowest-energy measurement outcomes and ignores the rest. 
As a result, a large amount of measurement information is wasted, which is undesirable 
for real-device implementations, where shot budgets are limited and data acquisition is expensive. 

\section{Design of our VQE algorithm}
\label{sec:design}
In this section, we detail the design of our VQE algorithm.
\subsection{Weighted CVaR}

In Reference~\cite{Barkoutsos2020ImprovingVariational}, 
the authors demonstrate that using the CVaR cost function leads to faster convergence 
and better solutions for the combinatorial optimization problems they tested, 
including portfolio optimization. Standard CVaR assumes that 
all outcomes within the lowest-energy $\alpha$-fraction 
contribute equally to the average, assigning them uniform weights. 
However, lower-energy outcomes are more directly related to high-quality solutions 
for combinatorial optimization problems. To better reflect this preference, 
a weighted sum may be more appropriate.
To enhance performance, we propose using a weighted CVaR as the cost function, 
where the weights can depend on either the energies or their ranks 
within the ordered sample. The WCVaR cost function is defined as:
\begin{align}
\mathrm{WCVaR}_{\alpha}(\boldsymbol E) &= \sum_{k=1}^{\lceil \alpha K\rceil} w_k E_{(k)},
\label{eq:wcvar}
\end{align}
where $w_k$ denotes the weight assigned to the $k$-th smallest energy $E_{(k)}$ 
in the ordered sample set, and the weights satisfy the normalization condition:
\begin{align}
\sum_{k=1}^{\lceil \alpha K\rceil} w_k &= 1. \label{eq:wsum}
\end{align}
The choice of weights plays a central role in WCVaR. We use the 
piecewise exponential weighting function, with details
given in Appendix~\ref{app:weights}.

\subsection{Covariance Matrix Adaptation Evolution Strategy}

While the mean sampled energy~\eqref{eq:vqecost} can be a poor cost function for VQE, 
both CVaR~\eqref{eq:cvar} and weighted CVaR~\eqref{eq:wcvar} offer significant improvements 
by focusing optimization on low-energy states. However, these cost functions 
are typically non-smooth and noisy, which can hinder gradient-based optimizers 
and make it difficult to locate the minimum reliably.
To address this challenge, we employ the Covariance Matrix Adaptation 
Evolution Strategy, a popular derivative-free optimization algorithm 
known for its effectiveness on nonlinear, non-convex, 
and noisy problems~\cite{Hansen2023CMAEvolutionStrategytutorial}. 
The core idea of CMA-ES is to iteratively improve a population 
of candidate solutions using an evolutionary strategy, 
gradually adapting the search distribution to converge toward the global optimum. 
Since CMA-ES does not require gradient information and 
is robust to noise, it is well-suited for optimizing the VQE cost function.

For benchmarking purposes, we also use the COBYLA optimizer, 
which has been reported as one of the most stable optimizers 
for VQE in Reference~\cite{Buonaiuto2023BestPracticesPortfolioOptimization}.

\subsection{VQE ansatz}

\begin{figure}[tbhp]
\raggedright
\includegraphics[width=0.9\linewidth]{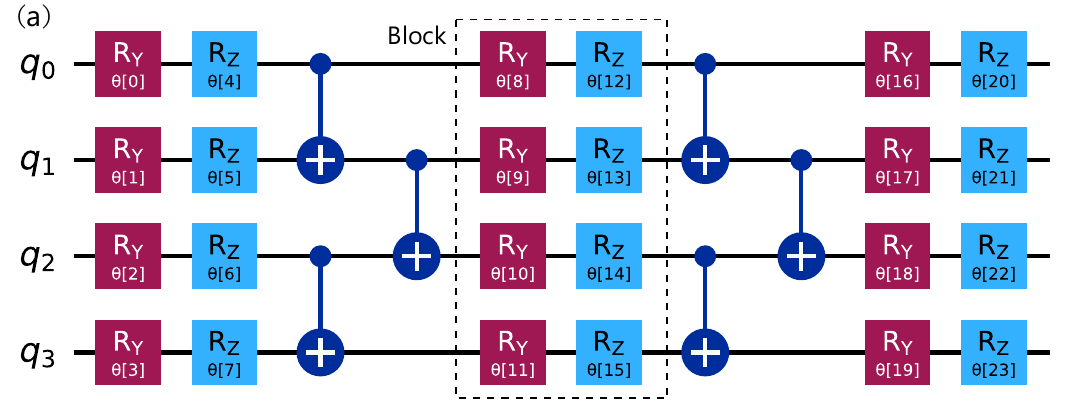}
\includegraphics[width=\linewidth]{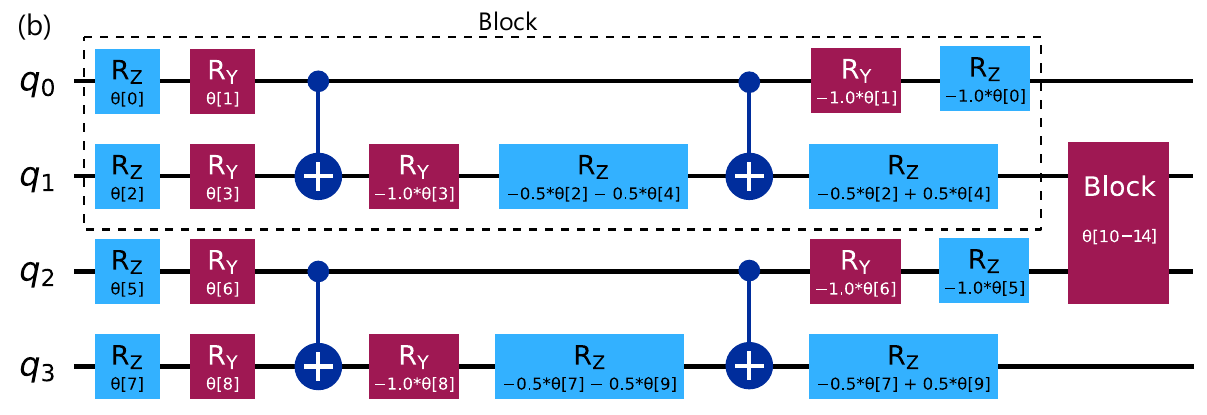}
\caption{The ansatzes for the VQE algorithm in this work. 
(a) A three-layer two-local ansatz. We use RY and RZ gates 
for the single-qubit rotation layers. Pairwise CNOT gates are used between layers. 
(b) A two-layer block ansatz. Each block contains two CNOT gates.
This quantum circuit diagram was plotted using Qiskit~\cite{Qiskit}.}
\label{fig:ansatz}
\end{figure}

Reference~\cite{Buonaiuto2023BestPracticesPortfolioOptimization} reports 
that the Pauli two-design ansatz performs best among the architectures evaluated. 
However, its use of random single-qubit rotation gates 
makes it difficult to reproduce and unsuitable for reliable benchmarking. 
Therefore, we adopt the two-local ansatz with RY and RZ gates as our first circuit structure. 
Specifically, we use the two-local ansatz with RY and RZ gates for the single-qubit rotation layers.
Additionally, motivated by the work of~\cite{Chivilikhin2020MOGVQEMultiobjective}, 
we design a second ansatz, the block ansatz, based on a general two-qubit building block. 
This block contains two CNOT gates and can implement arbitrary 
controlled-unitary operations~\cite{NielsenChuang2010QuantumComputationInformation}. 
In both ansatzes, entanglement is introduced through pairwise connections between adjacent qubits. 
The circuit diagrams for these two ansatzes are shown in Figure~\ref{fig:ansatz}.

\subsection{Wuyue platform}
We implemented our VQE algorithm using the Wuyue QuantumAI computing framework, 
a cloud-based platform~\cite{WuYue} for quantum computing. 
Numerical simulations were carried out using the simulator 
integrated into the Wuyue QuantumAI platform, 
where the main components of our implementation, 
including the CMA-ES optimizer and the WCVaR cost function, 
were provided within a unified cloud-based framework. 
Experiments were conducted using the quantum hardware resources 
connected to the same platform. Specifically, the hardware runs 
were executed on the Baihua superconducting quantum processor~\cite{Quafa}, 
which is developed and maintained by the Superconducting Quantum Computing 
Group of the Beijing Academy of Quantum Information Sciences.

\section{Results}
\label{sec:results}

We constructed a $12$-stock portfolio comprising $6$ Chinese A-shares 
and $6$ U.S. equities to enable a cross-market optimization analysis. 
The A-share constituents were selected to represent diverse sectors, 
including financial services, consumer staples, technology, and energy. 
In contrast, the U.S. holdings are primarily large-cap technology companies 
and other sector leaders. This composition provides diversified exposure 
across market capitalizations, industries, and geographic regions.
Daily closing prices for the calendar year 2024 were obtained 
using the Yahoo Finance API via the Python library yfinance~\cite{yahoo_finance, yfinance_package}. 
These price data were processed into daily return series 
and used to compute the covariance matrix, 
both of which serve as inputs for the subsequent portfolio optimization. 
The dataset is available for download at \cite{Dataset}.

\begin{figure}[tbhp]
\centering
\includegraphics[width=\linewidth]{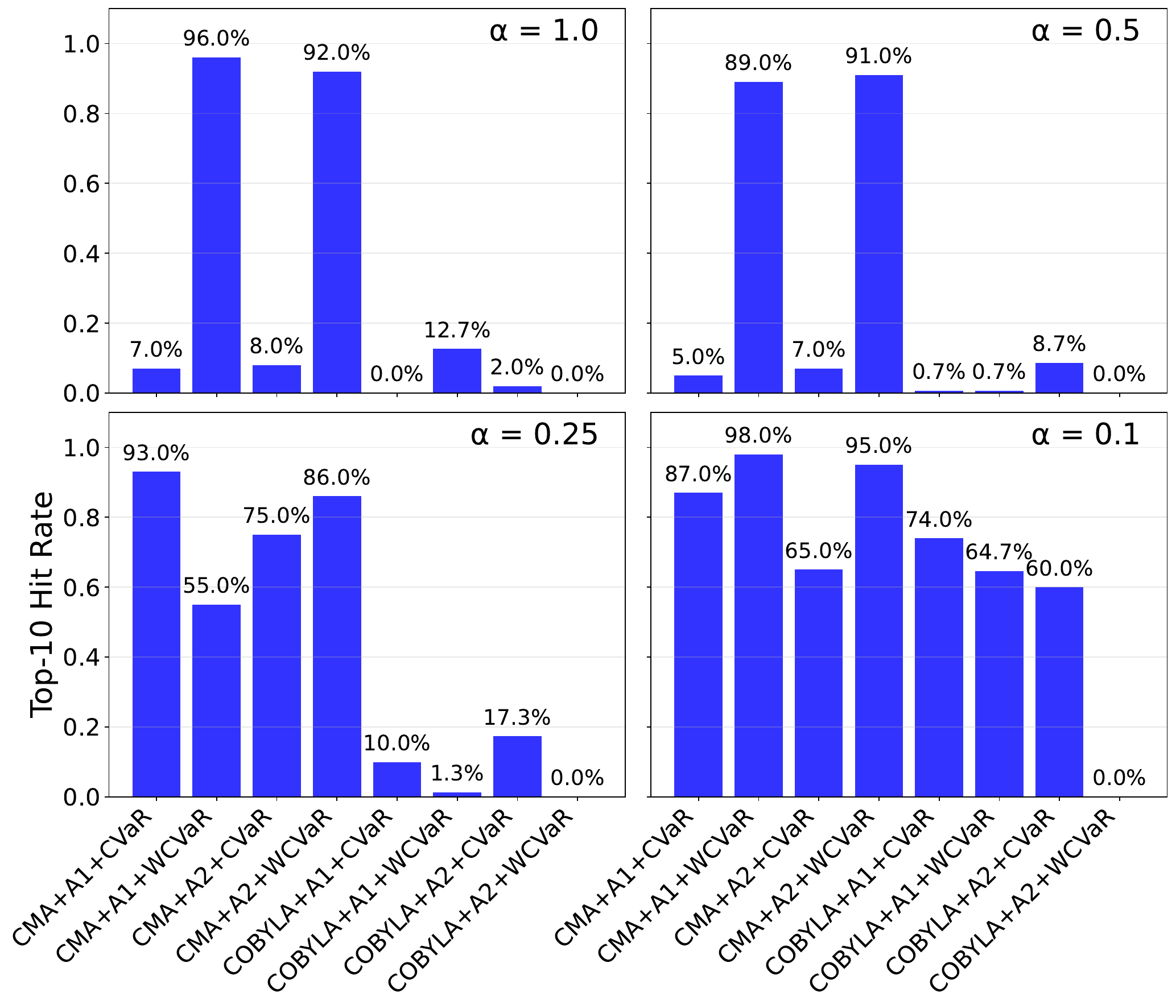}
\caption{Success rate during the optimization process 
for various combinations of ansatzes, optimizers, and cost functions. 
A1 and A2 denote the two-local ansatz and the block ansatz, respectively.}
\label{fig:accumulation}
\end{figure}

To evaluate the performance of the VQE algorithm, 
we assessed its success rate over the course of the optimization.
Specifically, in each iteration, the quantum state output by VQE 
was projected onto the computational basis. 
The algorithm was deemed successful in that iteration 
if the exact ground state (i.e., the optimal solution) 
was found among the top $10$ states ranked by their measurement probabilities. 
The overall performance was then quantified as the total number of successful iterations.

The results are presented in Fig.~\ref{fig:accumulation}, 
where the first $100$ iterations are shown for the CMA-ES optimizer 
and the first $150$ iterations for the COBYLA optimizer. 
It is rare for the VQE algorithm to find the optimal solution in some iteration 
and then fail to find it in the subsequent iterations. 
Therefore, Fig.~\ref{fig:accumulation} can also be interpreted 
as a measure of convergence speed: a higher success rate indicates faster convergence. 
There is no significant difference in performance 
between the two-local ansatz and the block ansatz. 
However, both the choice of optimizer and the cost function 
have a substantial impact on performance. The main findings are summarized as follows:
\begin{itemize}
\item The performance of the CVaR cost function improves as $\alpha$ decreases, 
whereas the performance of the weighted CVaR (WCVaR) cost function 
shows no significant dependence on $\alpha$. At $\alpha=1$ or $\alpha=0.5$, 
CVaR performs considerably worse than WCVaR. 
However, when $\alpha=0.25$ or $\alpha=0.1$, CVaR's performance, 
although slightly inferior, becomes comparable to that of WCVaR. 
The property that WCVaR's performance does not depend on $\alpha$ 
allows us to directly choose $\alpha=1$, 
thereby avoiding the impact of hyperparameter selection 
on the efficiency of the VQE algorithm and reducing the computational resources required on quantum devices.
\item On average, the CMA-ES optimizer outperforms the COBYLA optimizer. 
Further numerical analysis reveals that even with up to $300$ iterations, 
COBYLA still fails to achieve the performance level attained 
by CMA-ES within just $100$ iterations. Therefore, 
we conclude that CMA-ES is superior to COBYLA for this problem, 
and we restrict subsequent experiments to the CMA-ES optimizer.
\item The COBYLA optimizer appears to be more compatible with 
the CVaR cost function, particularly at $\alpha=0.25$, 
while the CMA-ES optimizer demonstrates greater 
compatibility with the WCVaR cost function. 
This may suggest that the choices of optimizer, 
cost function, and even ansatz should be considered holistically, 
as there is no single optimal choice for any one aspect in isolation.
\end{itemize}

\begin{figure}[tbhp]
\centering
\includegraphics[width=\linewidth]{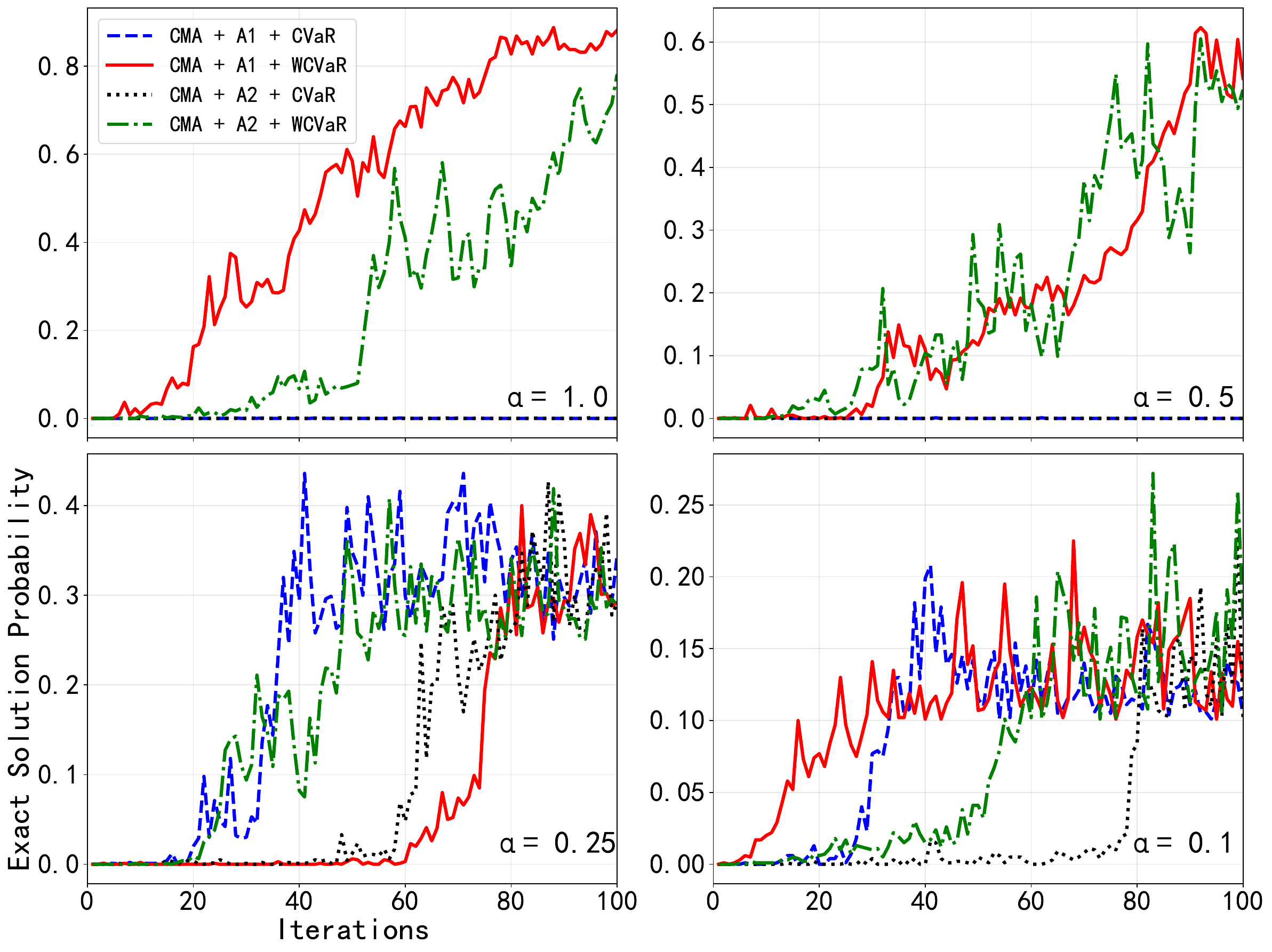}
\caption{The probability of sampling the
optimal solution for various combinations of ansatzes, optimizers, and cost functions. 
A1 and A2 denote the two-local ansatz and the block ansatz. }
\label{fig:pvsi}
\end{figure}

In Fig.~\ref{fig:pvsi}, we present the probability of sampling 
the optimal solution (i.e., the overlap between the VQE variational state and the ground state) 
as a function of the number of iterations. 
For consistency with previous results, 
only the first $100$ iterations are shown for the CMA-ES optimizer. 
The observed trends are consistent with those in Fig.~\ref{fig:accumulation}.



The above numerical results suggest that, although the $\alpha$-fraction 
is included in the WCVaR cost function, the performance of the VQE algorithm 
is only weakly dependent on the choice of $\alpha$. 
The role of $\alpha$ is to provide a stepwise weighting 
over the basis states in the variational wave function. 
In contrast, within the WCVaR framework, the weight $w_k$ plays a more dominant role. 
Therefore, the choice of $\alpha$ has only a limited effect on the performance 
of the VQE algorithm, and we may simply set $\alpha = 1$ in practice.
This choice ensures that all measurement outcomes from the quantum device contribute 
to the evaluation of the cost function, thereby making full use of the available readout data 
and improving measurement efficiency on real hardware.
In the following, we present the performance of the VQE algorithm on quantum hardware. 
We use CMA-ES as the optimizer and WCVaR as the cost function 
with the best-performing weights $w_k$. 
Details of the weight selection are given in Appendix~\ref{app:weights}. 
The VQE algorithm is executed for a six-stock portfolio comprising only 
Chinese A-shares on Baihua, 
a superconducting quantum computer with more than one hundred qubits.

Before executing the circuits on the quantum device, 
we calibrated the hardware parameters, including readout errors 
and single- and two-qubit gate errors, and selected the six best-connected qubits 
with the highest quality. The gate errors were characterized using randomized benchmarking, 
with interleaved randomized benchmarking used to estimate 
the fidelities of specific gates~\cite{Magesan2011RB,Magesan2012IRB}. 
The quality of these qubits is summarized in Table~\ref{tab:qubit_params}.

\begin{table}[h]
    \centering
    \caption{Calibration data for the six qubits selected on the Baihua superconducting quantum processor.}
    \label{tab:qubit_params}
    
    \begin{tabular}{lcccccc}
        \toprule
        & \multicolumn{6}{c}{Qubit} \\
        \cmidrule(lr){2-7}
        Parameter & 32 & 33 & 34 & 21 & 22 & 23 \\
        \midrule
        T1 ($\mu$s)           & 49.3 & 76.7 & 31.3 & 60.9 & 78.4 & 107.4 \\
        T2 ($\mu$s)           & 2.8 & 34.2 & 13.1 & 6.3 & 40.3 & 58.4 \\
        f (GHz)               & 4.4 & 4.0 & 4.4 & 4.0  & 4.4  & 4.1  \\
        Gate fidelity    & 0.971  & 0.999  & 0.998  & 0.998  & 0.999  & 0.999 \\
        Readout F0  & 0.966  & 0.978  & 0.985  & 0.986  & 0.972  & 0.986  \\
        Readout F1  & 0.898 & 0.977  & 0.982 & 0.976  & 0.981  & 0.975   \\       
        \bottomrule
    \end{tabular}

    \vspace{0.4em}

    \begin{tabular}{lc}
        \toprule
        Qubit pair & Multiqubit fidelity \\
        \midrule
        (32,33) & 0.980 \\
        (33,34) & 0.957 \\
        (34,21) & 0.952 \\
        (21,22) & 0.982 \\
        (22,23) & 0.978 \\
        \bottomrule
    \end{tabular}
\end{table}

\begin{figure}[tbhp]
\centering
\includegraphics[width=\linewidth]{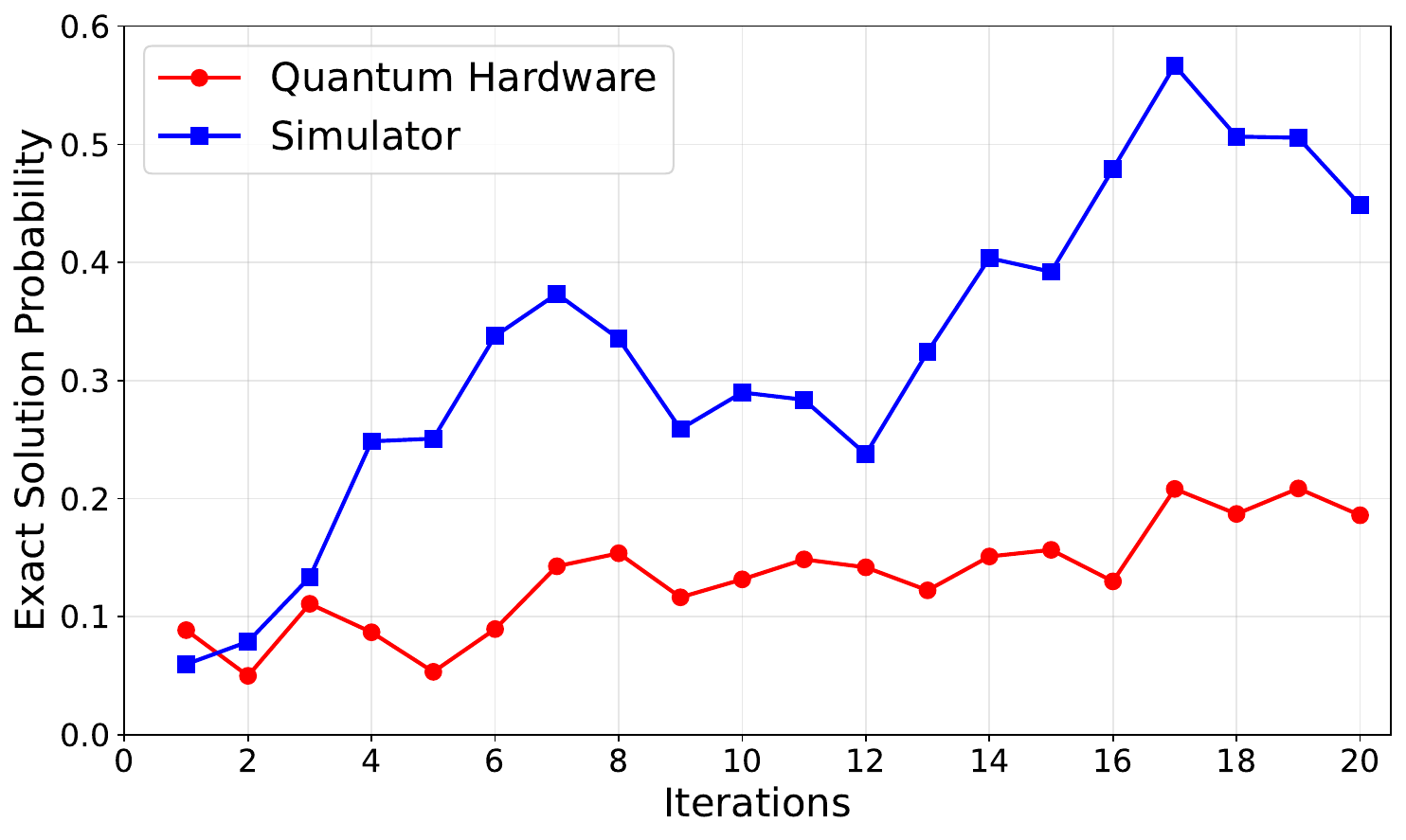}
\caption{Probability of sampling the exact ground state during VQE optimization 
on the Baihua superconducting quantum processor and in the corresponding Qiskit simulation. 
The algorithm is applied to a six-stock portfolio instance using the WCVaR cost function 
and the CMA-ES optimizer.}
\label{fig:real}
\end{figure}


In Fig.~\ref{fig:real}, we show the performance of the VQE algorithm 
on the Baihua quantum device. Within 20 iterations, the algorithm prepares a state 
with a $20\%$ overlap probability with the exact ground state of the Ising model 
corresponding to the portfolio optimization problem.
No error mitigation techniques are applied except measurement error correction.
Although the performance is not as strong as in the simulation results, 
the experiment still demonstrates a clear concentration of probability 
on the optimal solution. 
This provides preliminary evidence that the proposed WCVaR-based VQE framework 
is effective on real quantum hardware and can produce useful solutions 
for practical portfolio optimization instances.

\section{Conclusions}
\label{sec:conclusions}

In this work, we investigated the portfolio optimization problem 
using the Variational Quantum Eigensolver (VQE) algorithm. 
We formulated the problem as a QUBO model and mapped it to an Ising Hamiltonian, 
making it suitable for implementation on gate-based quantum devices. 
To improve the optimization performance, we proposed a Weighted Conditional Value-at-Risk 
(WCVaR) cost function and combined it with the Covariance Matrix Adaptation Evolution 
Strategy (CMA-ES) optimizer.

Using a 12-stock dataset consisting of Chinese A-shares and U.S. equities, 
we systematically compared different ansatzes, cost functions, and optimizers 
on the Wuyue QuantumAI simulator. 
The simulation results show that the choice of cost function and optimizer 
has a significant impact on convergence behavior and solution quality. 
Among the configurations tested, the combination of WCVaR and CMA-ES achieved 
the best overall performance. 
In particular, WCVaR shows weak dependence on the retained fraction $\alpha$, 
which allows us to use all measurement outcomes in the cost evaluation 
and thereby improves measurement efficiency.

We further implemented the proposed method on the Baihua superconducting quantum processor 
for a six-stock portfolio instance composed of Chinese A-shares. 
Without applying error mitigation techniques other than measurement error correction, 
the algorithm prepared a state with a $20\%$ overlap probability with the exact ground state 
within 20 iterations. 
Although the hardware performance remains below the noiseless simulation results, 
the experiment demonstrates a clear concentration of probability on the optimal solution. 
These results provide preliminary evidence that the proposed WCVaR-based VQE framework 
is effective on real quantum hardware and can produce useful solutions 
for practical portfolio optimization instances.

Future work will focus on scaling the problem size, 
incorporating more realistic portfolio constraints, 
and improving robustness against hardware noise. 
It would also be valuable to combine the proposed cost-function design 
with advanced error mitigation techniques and hardware-aware ansatz optimization 
to further improve performance on NISQ devices.

\begin{acknowledgments}
We acknowledge the support from 
Basic Research for Application Program of China Mobile (No. R251166S).
\end{acknowledgments}

\appendix

\section{Weights in WCVaR cost function}
\label{app:weights}

In this section, we describe the weighting schemes employed in the WCVaR cost function. 
We assume the sampled energies are sorted in ascending order, 
forming an ordered set $E = (E_{(1)}, E_{(2)}, \ldots, E_{(K)})$, 
where $E_{(k)}$ denotes the $k$-th smallest energy value. 
For the exponential weighting schemes below, we present the unnormalized weights; 
in implementation, they are normalized according to Eq.~\eqref{eq:wsum}.

The standard Conditional Value-at-Risk (CVaR) can be viewed as a special case of WCVaR, 
where equal weights are assigned to the $\lceil \alpha K\rceil$ 
smallest values in the sample set, and zero weight to all others:
\begin{align}
w_k &= \begin{cases}
1/\lceil \alpha K\rceil & \text{if } k \leq \lceil \alpha K\rceil \\
0 & \text{otherwise}
\end{cases}
\end{align}

The exponential weighting function is well-known to researchers 
in both physics and computer science, as it arises naturally in Boltzmann's law 
in statistical physics and the softmax function in machine learning and optimization. 
In this work, we employ three distinct variants of the exponential weighting function 
to explore their impact on the WCVaR cost function.

The first variant is the energy-based exponential weighting:
\begin{align}
w_k &= \exp\left[-\beta(E_{(k)} - E_0)\right],
\end{align}
where $E_0 = E_{(1)}$ denotes the minimum energy value observed in the sample set, 
and $\beta$ is an inverse temperature parameter 
that controls the concentration of weights on low-energy states.

The second variant is the rank-based exponential weighting:
\begin{align}
w_k &= \exp\left(-\beta k\right),
\end{align}
where $\beta$ is again an inverse temperature parameter. 
This form assigns weights based solely on the rank $k$ of the energy value, 
independent of its absolute magnitude.

The third variant is a piecewise exponential weighting function, 
designed to allow different decay rates across segments of the ordered sample:
\begin{align}
w_k &= \begin{cases}
\exp\left(-\beta_1 k\right) & \text{if } k < N_1, \\
\exp\left(-\beta_2 (k-N_1)\right) \cdot w_{N_1-1} & \text{if } N_1 \le k < N_2, \\
\exp\left(-\beta_3 (k-N_2)\right) \cdot w_{N_2-1} & \text{if } k \ge N_2.
\end{cases}
\end{align}
Here, $N_1$ and $N_2$ (with $N_1<N_2$) are positive integers 
defining the transition points between segments, and $\beta_1,\beta_2,\beta_3>0$ 
are positive decay parameters. The continuity of the weights at $k=N_1$ and $k=N_2$ 
is ensured by multiplying the exponential in each subsequent segment 
by the weight value at the end of the previous segment (e.g., $w_{N_1-1}$ and $w_{N_2-1}$).

\begin{figure}[tbhp]
\centering
\includegraphics[width=\linewidth]{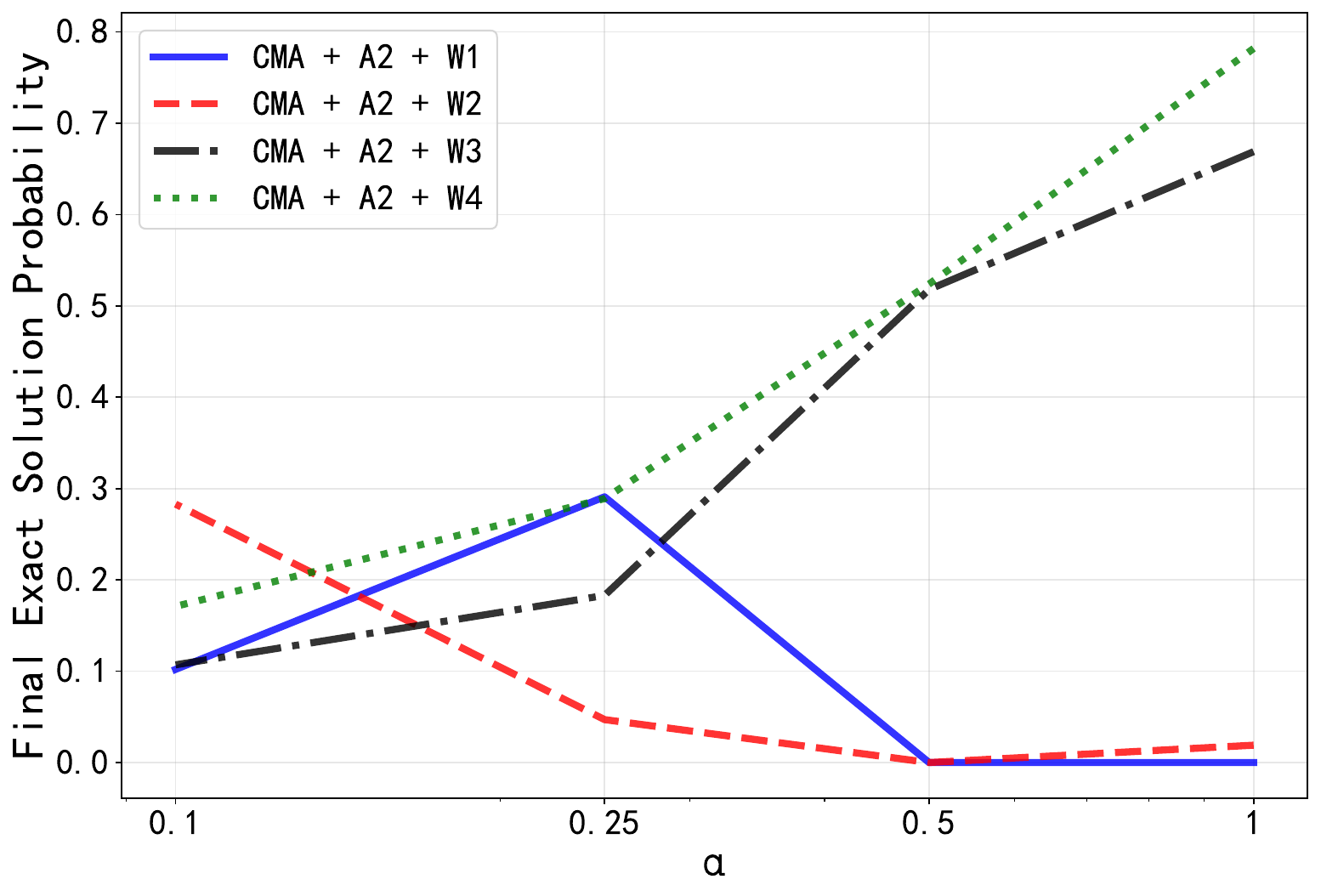}
\caption{Probability of sampling the optimal solution under different weighting schemes. The blue solid line (W1), red dashed line (W2), black dash-dotted line (W3), and green dotted line (W4) denote the CVaR, energy-based exponential weighting, rank-based exponential weighting, and piecewise exponential weighting cost functions, respectively.}
\label{fig:weights}
\end{figure}

In Fig.~\ref{fig:weights}, we present the probability of 
sampling the optimal solution under four different weighting schemes. 
CVaR performs well when $\alpha$ is small but poorly when $\alpha$ is large. 
The energy-based exponential weighting scheme yields the worst performance; 
therefore, its piecewise variant is not considered further. 
In contrast, the rank-based exponential weighting performs well, 
particularly when $\alpha$ is large. 
The piecewise exponential weighting function outperforms the rank-based version, 
although the improvement is not significant. 
Consequently, we adopt the piecewise exponential weighting function 
in the main part of this work. However, if minimizing the number of hyperparameters 
is a priority, the rank-based weighting function 
remains a viable and effective alternative.

\bibliography{reference}

\end{document}